# Electron Emissions and Hot spots in Dual-Phase LXe TPCs

J. Va'vra

SLAC, Stanford University, CA94309, U.S.A.
e-mail: jjv@slac.stanford.edu

## Abstract

**Persistent photon and single-electron emissions - in the form of "electron trains" and localized "hot spots" - have been observed in multiple dual-phase liquid xenon (LXe) time projection chambers (TPCs), often persisting long after ionizing events. We show that these phenomena are naturally explained by photon-triggered single-electron emission from resistive mixed-oxide films on stainless-steel wires, which behave as leaky capacitors with long RC time constants at LXe temperature. Positive ions landing on these oxides can further enhance local fields and drive Malter-like electron emission. We outline the materials physics ($Cr_2O_3$/$Cr_2O_{3-x}$/$Cr(OH)_3$ mosaics), quantify expected time scales (~1 second under illumination), and demonstrate how small damaged regions with enhanced QE can produce persistent hot spots.**

## Introduction

Table 1 summarizes observations from several dual-phase LXe TPCs—including LUX, XENON1T, XENONnT, LZ, and PandaX—reporting persistent single-electron (SE) signals and localized emission regions, so-called "hot spots." One common problem is that photon and electron activity lasts about a second after each large signal – S2 (electroluminescence) or S1 (prompt scintillation). This was a puzzle not explained ever since the dual-phase LXe TPC detectors started. We propose and quantify a mechanism that can account for the observed ~1 second tails and localized hot spots.

LZ detector extraction region is shown on Fig.1 including present anode and grid running voltages. Gas section is 8 mm thick and liquid section is 5 mm thick. Extracted electrons from LXe enter gas section and are subject to electric field of ~6.2 kV/cm. They excite Xe atoms over path of 8 mm producing about 400 photons per each electron. Example of electron train (red dots) is shown on Fig.2, where S2 signal triggers a train of single electron [1]. An example of hot spots observed in LZ TPC is shown on Fig.3 [2], including train of pulses.

**Table 1**

| Experiment | Observation | Term used | Comment |
|---|---|---|---|
| LZ | Occasional persistent SE emission post-muon | "Delayed SE", "Hot spots", "electron trains" | Interpreted as low-level background |
| Xenon1T | Occasional electron trains, localized emission | "Hot spots" | Seen during low background runs |
| PandaX | Time-dependent SE backgrounds | "After-pulses" | Not clearly linked to known triggers |

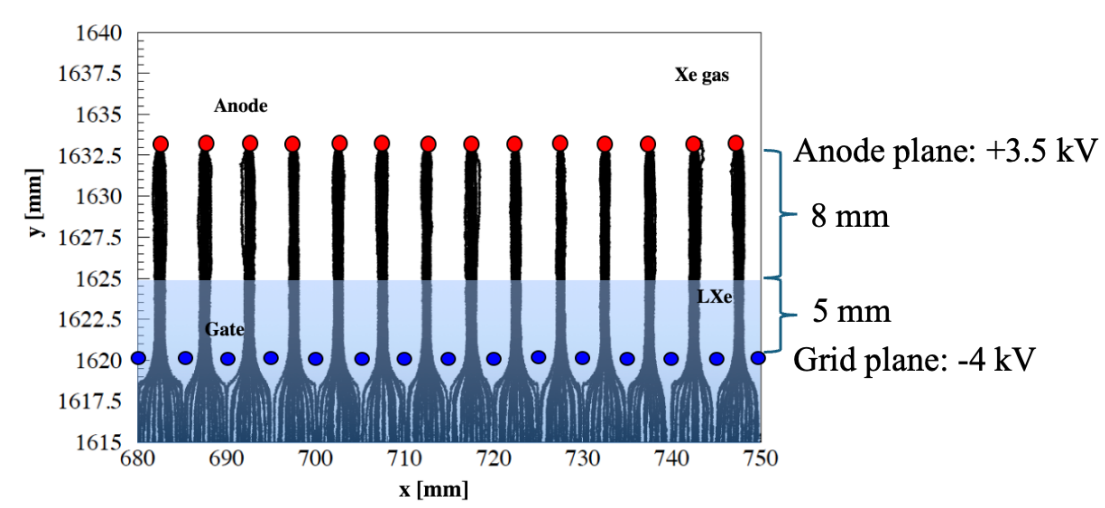

**Fig. 1** LZ design of the extraction region in dual-phase LXe TPC, including the present running voltages. Gas gap 8 mm, field ≈ 6.2 kV/cm, S2 light yield ≈ 400 γ/e⁻.

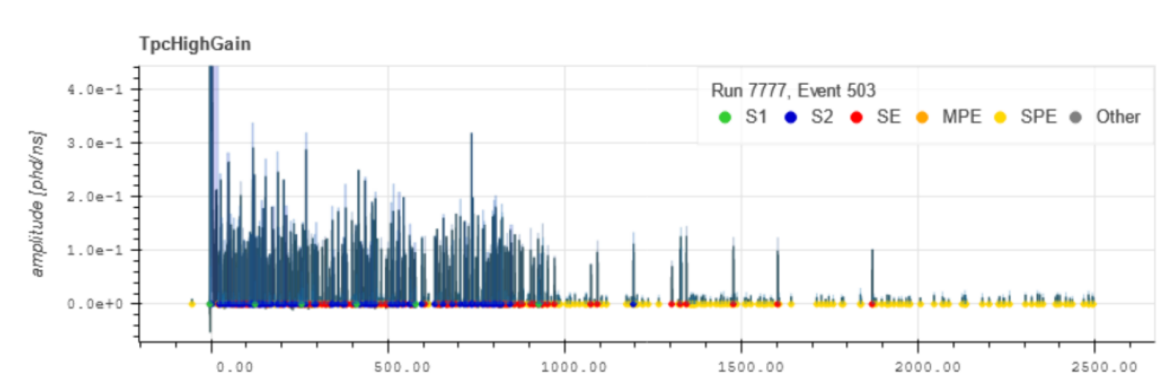

**Fig. 2** S2 pulse triggering single-electron (SE) train lasting about a second [1].

This work supported by the Department of Energy, US contract DEAC02-76SF00515.

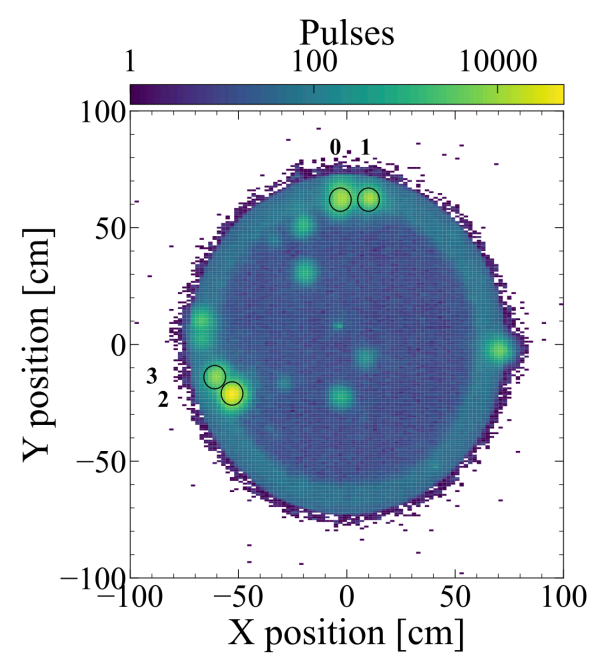

**Fig. 3** Hot spots observed in LZ [2].

## 1. Chemistry of passivation of stainless-steel wires

Historically, passivation of 304 stainless-steel was done typically with nitric acid (20–50 % $HNO_3$ at ~50 °C for 30–60 min). The nitric acid dissolves free iron and iron-rich surface inclusions (neutral Fe atom, FeO, $Fe_2O_3$, $Fe_3O_4$). Chromium and nickel are not attacked appreciably, so the surface becomes **Cr-enriched**. After rinsing and drying in air, the freshly exposed Cr metal oxidizes spontaneously within minutes, forming a continuous $Cr_2O_3$ film typically 2 nm thick on top of (Fe/Cr/Ni) metal [3]. The passivation is making the wire smoother, which suppresses local field enhancements. The 2 nm layer of $Cr_2O_3$ suppresses surface photoemission because photoelectrons generated by VUV photons must tunnel through this ~2 nm oxide. As a result, the photoelectric yield QE drops.

LZ experiment has decided to passivate s.s. wires in citric acid [4]. The chemistry of oxide layer is more complicated for this treatment [5], [6]. In addition to $Cr_2O_3$ oxide, the process produces $Cr(OH)_3/Cr_2O_{3-x}$ oxides (variable x indicates how many oxygen are missing; x=0 corresponds to $Cr_2O_3$ (x is 0.01-0.1 typically). The oxide film is typically ~2 nm thick overall. Near Ni/Fe/Cr metal interface, the oxide is dominated by $Cr_2O_3$ oxide, which is dense and crystalline. Above, there is a mosaic of $Cr_2O_3$ and $Cr_2O_{3-x}$. Top layer is $Cr(OH)_3/Cr_2O_{3-x}$ oxide mix. Such mosaic of three oxides has (a) lower resistivity by a factor of 10x or even more, because $Cr(OH)_3$ is more hydrated, and (b) higher dielectric constant (~2x higher than $Cr_2O_3$).

## 2. Band structure of stainless-steel wire after passivation

Although real theory of oxides is complicated [7], it is usually expressed in terms of band structure just like semiconductors. Figure 4 shows the band structure of oxide on stainless-steel after passivation. To explain this in simple way, we use equivalent model. Atoms in solid oxide (say, $Cr_2O_3$) have their atomic orbitals overlapping. Instead of having discrete electron levels (1s, 2s, 2p, etc.) like an isolated atom, this overlap produces **band structure** - broad ranges of allowed energy levels for electrons: (a) **Valence Band**: Its states are completely or partially filled with electrons, electrons cannot freely move. (b) **Conduction Band**: It states are empty or partially filled; electrons in this band mobile and transfer to LXe. (c) **Band Gap**: The energy range between the valence band and the conduction band where no band states exist; defect states may lie inside.

Band gap energy Eg is the minimum energy difference between the top of the valence band and the bottom of the conduction band in a material, which an electron must gain to move from valence band to LXe. In semiconductors, this band gap energy Eg is ~ 2 eV (typically 1-3 eV), in insulators this gap is ≥3 eV; for example, it is 3.0 eV in $Cr_2O_3$ oxide, 0 eV for metals and 9 eV in quartz.

In **conductors**, there is no gap (Eg ~0eV). Emission is set by work function and surface fields. Metal is allowing electrons to move freely. **Insulators** have a large band gap, making it very difficult for electrons to jump from the valence band to the conduction band unless the absorb energy from VUV photons. **Semiconductors** have a moderate band gap, allowing a small number of electrons to be excited into the conduction band under certain conditions.

In Fig.4, we see that next to (Ni/Fe/Cr) metal surface there is Schottky barrier and Fermi level,[1] both within valence band; to get electron into oxide, electrons have to tunnel through Schottky barrier into the oxide; the barrier also suppresses return to metal. Closest to metallic surface is crystalline $Cr_2O_3$ oxide layer. Above it there is a mosaic of $Cr_2O_3$/ $Cr_2O_{3-x}$ oxide mix and layer above it is dominated by $Cr(OH)_3/Cr_2O_{3-x}$ oxides. Overall, the band gap is full of defects, which are traps for photoelectrons. Oxide resistivity is lower at room temperature, which is explained that electrons move slowly in the direction of field by action of thermal energy from one trap to another, slowly discharging capacitor; however, resistivity is much higher at LXe temperature, because electrons are trapped much longer and thermal energy does not help as much.

In real metal, the photoemission is prompt, proportional to field. In oxides, the photoemission is suppressed, because electrons have to tunnel through the oxide (across the band gap and/or defect states). Band gap accumulates photoelectrons in its defects by action of VUV light, thus charging the capacitor, which oxide layer represents.

[1] Fermi level controlls removal of electrons from metal.

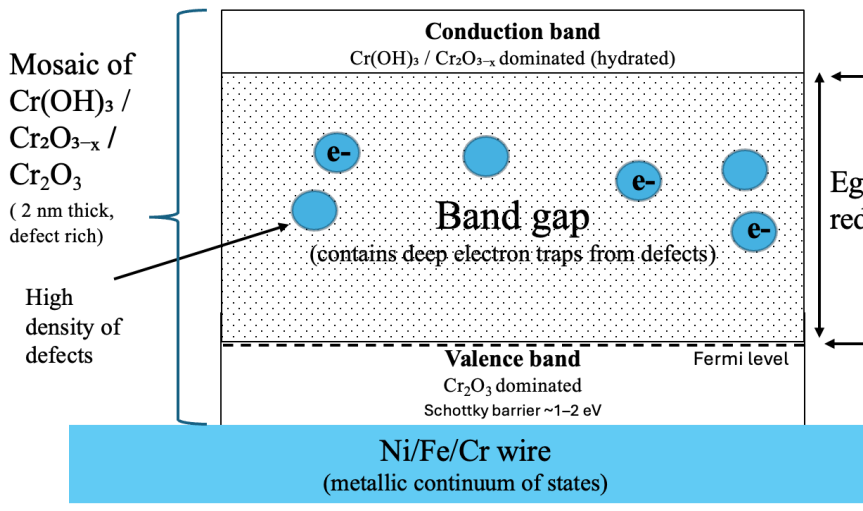

**Fig. 4** Schematic band structure of the mixed-oxide passivation layer on stainless-steel LXe TPC cathode wires. The citric-acid-treated surface forms a ~2 nm defect-rich mosaic of $Cr(OH)_3$ / $Cr_2O_{3-x}$ / $Cr_2O_3$. The band gap (Eg ~3 eV) contains deep electron trap states that can store charge for long times at LXe temperature. VUV photons can release these trapped electrons one-by-one into the liquid, producing delayed single-electron "train" signals.

## 3. Chromium oxide $Cr_2O_3$ resistivity

Figure 5a shows that $Cr_2O_3$ resistivity increases rapidly with reducing temperature [7]. Figure 5b shows real data from 1954 [8]. I extrapolate these data to LXe temperature - see Figures 6a&b. One concludes that the volume resistivity of **~5x10$^{15}$ Ω.cm** is possible at LXe temperature.

If such high values of oxide resistivity can be reached, this problem deserves attention. A resistive film creates a "leaky capacitor", which, if charged (by VUV photon flux), it will take a longtime constant RC to discharge, if operated at LXe temperature. This time constant is $\tau = RC = R\, \varepsilon_0\, \varepsilon_r\, A/d = \varepsilon_0\, \varepsilon_r\, \rho_{film}$, where $\varepsilon_0$ =8.852x10$^{-12}$ F/m, $\varepsilon_r$ = 10-12 for $Cr_2O_3$ oxide, R is resistance, A is area and d is thickness of resistive film with volume resistivity of $\rho_{film}$. At room temperature $Cr_2O_3$ oxide resistivity is $\rho$ ~10$^9$ Ω·cm. The corresponding RC time constant is $\tau$ ~3.5 milliseconds, which is acceptable because room temperature drift detectors with this wire worked well. However, at LXe temperature: $\rho_{film}$~5x10$^{15}$ Ω·cm yielding time discharging constant $\tau$~1.5 hours! This sharp contrast illustrates how oxide films become charge-retaining at cryogenic temperatures.

(a)

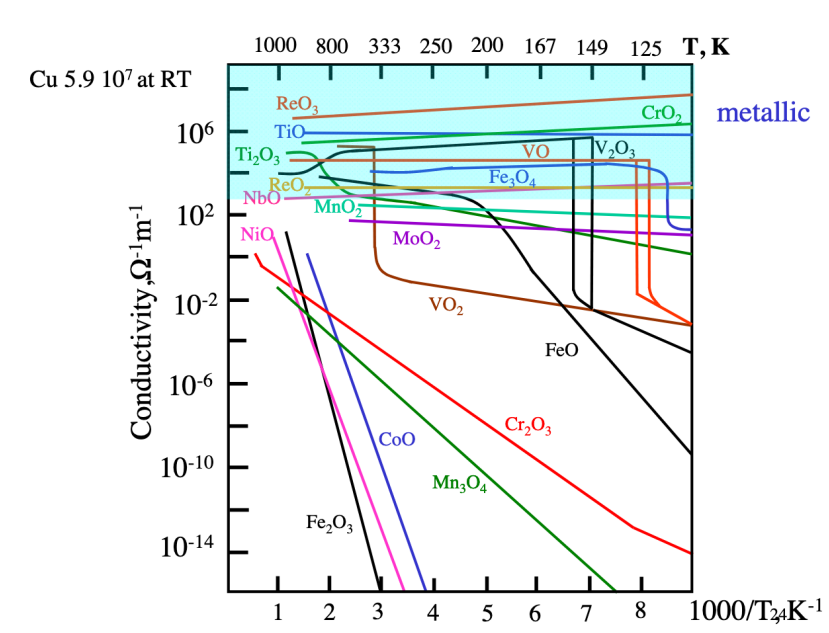

(b)

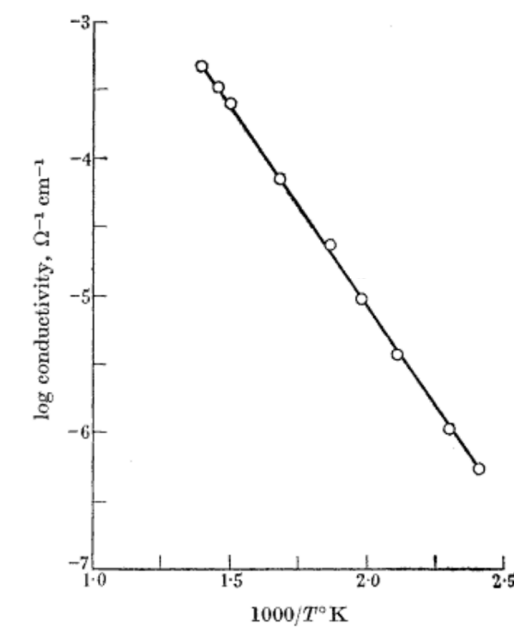

**Fig. 5** (a) This graph shows that some metal oxides belong to metallic group with very small resistivity (shown in light blue). However, other oxides have high resistivity especially when cooled to very low temperature. For example, chromium oxide $Cr_2O_3$ belongs to high resistivity oxides at LXe temperature [7]. Figure seems to describe various oxides schematically. (b) A real measurement of conductivity of pure $Cr_2O_3$ in oxygen [8].

(a)

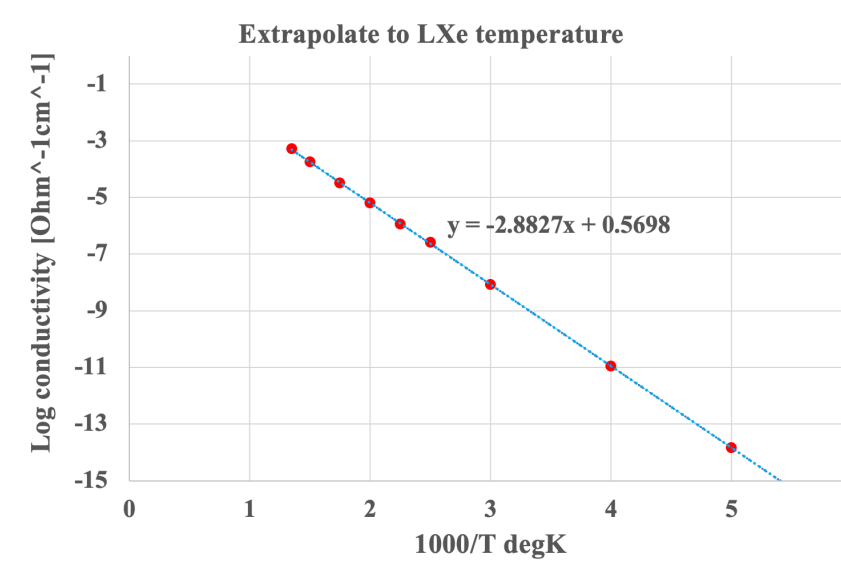

(b)

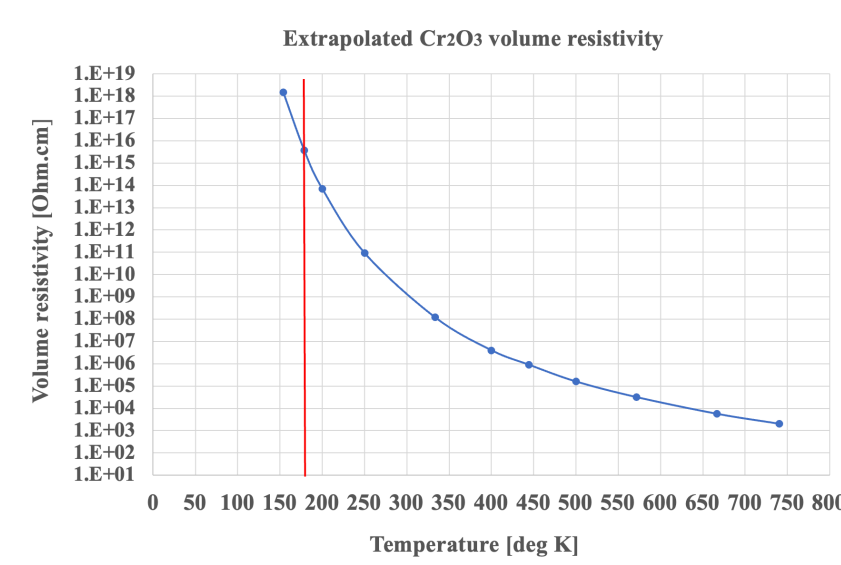

**Fig. 6** (a) My extrapolation based on available data in Fig.5b. (b) Extrapolation to LXe temperature.

## 4. Photoelectric effect on passivated wires

Passivation generally suppresses QE of photoeffect through two major effects: (a) making the wire smoother, suppressing local field enhancements, (b) a few nm oxide film suppresses the photoemission, because photoelectrons generated by VUV photons must tunnel through this oxide.

The most recent measurement of QE of several metals was done by Kazama [9]. They measured QE ≈ 3.2 × 10$^{-4}$ electrons/photon for SS304 stainless steel wire (Fe/Cr/Ni metal), QE ≈ 2.5 × 10$^{-4}$ for PT wire and QE ≈ 7 × 10$^{-5}$ for $MgF_2$-coated aluminum wire, all using **178 nm** and in LXe at 173 K. However, they did not passivate their s.s. wires and therefore their s.s. wires were only partially oxidized because of handling wires in air; therefore, QE was only partially suppressed. If they would passivate s.s. in nitric acid and develop $Cr_2O_3$ oxide, the QE would further decrease by a factor 3-6 compared to non-passivated s.s. wire [9],[10].

Table 2 in Appendix #1 shows a QE list of many wires. It is clear that QE depends strongly on how the wire is prepared.

The s.s. wire is formed from Fe/Cr/Ni alloy. If this alloy is <u>not oxidized</u>, its QE is about ≈ $10^{-3}$-$10^{-2}$ $e^-$/photon at 178 nm [11], roughly comparable to ultra-clean Cu or Al.

## 5. Oxide damage during cooling TPC wires to LXe temperatures

During wire weaving, any surface abrasion is rapidly "healed" by re-oxidation. However, once the mesh is assembled and operated in the oxygen-free xenon environment, subsequent mechanical motion — due to thermal contraction during cool-down or electrostatic flexing under high voltage - can locally rupture brittle oxide. **<u>At this point, "healing" is impossible because there is no oxygen</u>**. One ends up with damaged spots.

$Cr_2O_3$ behaves like classical brittle ceramic (like $Al_2O_3$, sapphire, and ruby), easy to crack of flake. $Cr_2O_{3-x}$ still behaves as ceramic, though a bit more compliant. The $Cr(OH)_3$ oxide is not a crystalline ceramic, but at LXe temperature it behaves glassy/brittle, prone to cracking. Thermal expansion coefficients:

a) $Cr_2O_3$ (≈ $8\times10^{-6}$ $K^{-1}$) << Fe/Cr/Ni stainless steel metal (≈ 16–$18\times10^{-6}$ $K^{-1}$).

b) $Cr_2O_{3-x}$ (≈ 8-$10\times10^{-6}$ $K^{-1}$)

c) $Cr(OH)_3$ contracts irregularly as one cools it, becomes more brittle and can flake.

As a result, there will be thousands of micro-cracks exposing Fe/Cr/Ni bare wire when such s.s. wire is cooled to LXe temperature of -97°C. Figure 7 describes the picture schematically. Notice that there could be a lot of sharp points due to a different shrinkage of oxides relative to metal. QE may increase locally by a factor of 10 – 200x. Sharp points may enhance this factor further; a factor 1000x is possible with field enhancement.

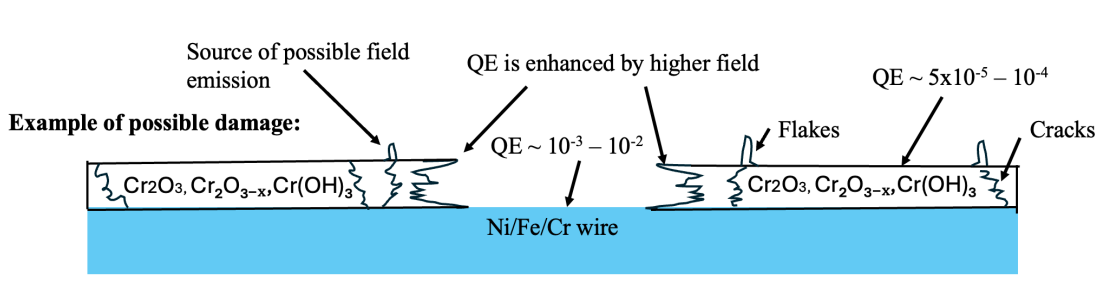

**Fig. 7** Picture shows damaged oxide when passivated s.s. cools to LXe temperature. There are cracks, flakes, sharp points, gaps because these oxides are brittle and have ~2 times lower coefficient of thermal expansion compared to Ni/Fe/Cr bare alloy.

SEM (scanning electron microscopy) might reveal larger oxide cracks and flake detachments, but the 1–3 nm mixed-oxide layer and its sub-nanometer defect structure are below SEM resolution and may require TEM (Transmission Electron Microscope).

## 6. Electron and photon emissions

There are three types of electron and photon emissions: (a) oxide -mediated delayed emissions, (b) emissions due to oxide damage of s.s. wires and (c) ion-induced emission.

### a) Oxide-mediated delayed emission and electron trains

For the $Cr_2O_3/Cr(OH)_3/Cr_2O_{3-x}$ mosaic of oxides resistivity is lower; if we assume $\rho_{film}$ ~$5\times10^{13}$ Ω·cm and 2x larger $\varepsilon_r$, we ontain τ~36 minutes. When charged, oxide capacitor takes long tim to discharge. However, if the oxide is irradiated by photon flux during S2 or S1, the discharge time is reduced to τ ~1 second, as shown on Figures 2&8; this would correspond to ρ ~$\tau/(\varepsilon_0\varepsilon_r)$ ~$5.6\times10^{11}$ Ω·cm during illumination (for $\varepsilon_r$ ~20).

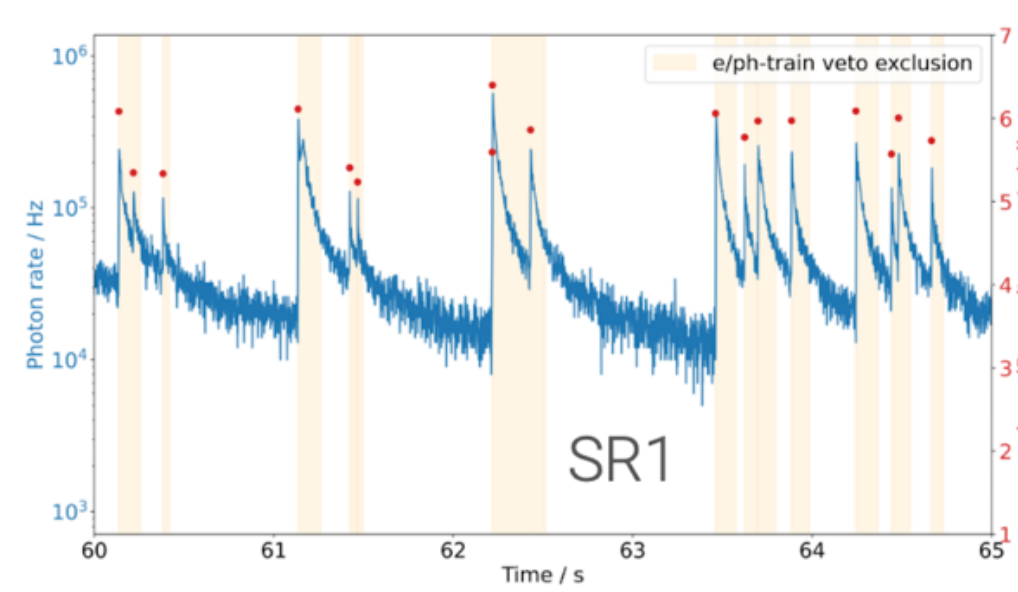

**Fig. 8** Typical discharge times after each S2 or S1 observed in LZ [1].

Figure 9 shows the process of formation of electron trains. Assuming that the oxide is charged with electrons from previous VUV activity in TPC. When a large S2 occurs, large number of VUV photons are kicking out electrons from the band gap into LXe, one by one, thus forming **<u>electron trains</u>**. Bare Fe/Cr/Ni wire will have even higher probability to produce the train, especially if QE is field enhanced.

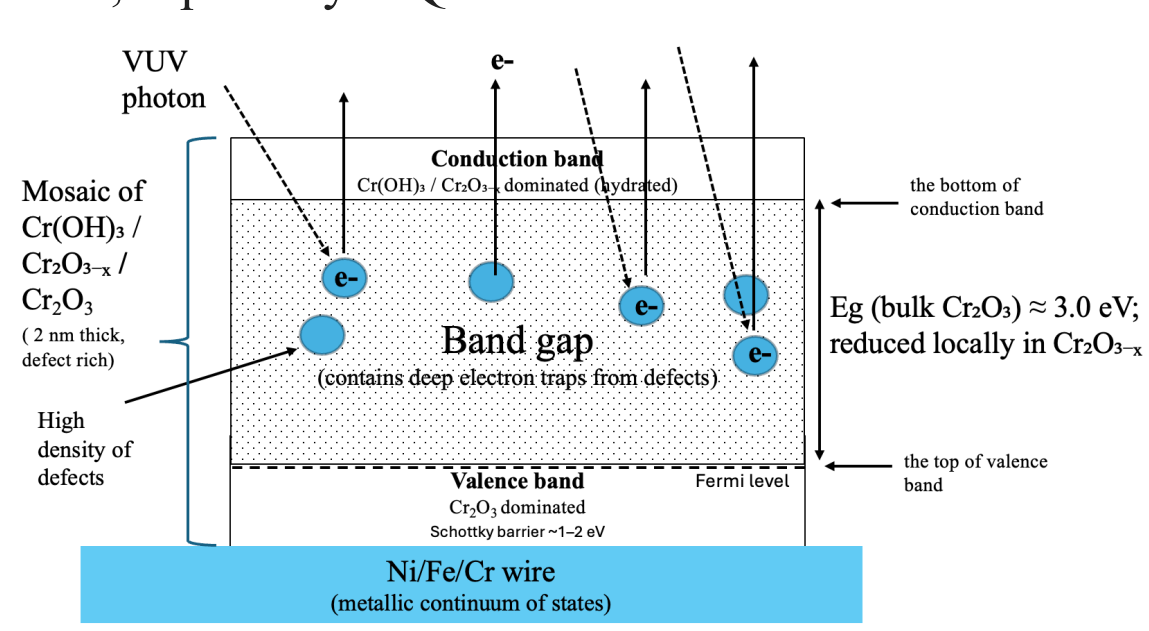

**Fig. 9** VUV Photons from S2 will knock out electrons from the oxide capacitor into LXe, one by one, thus creating electron trains similar to the one shown on Fig. 2.

## b) Formation of "hot spots" due to damage of wire oxides

Some hot spots, as shown on Fig.3, appear on their own, sometimes randomly. I explain this with help of Fig.8: due to a different shrinkage of oxide and metal during cooling, oxide is full of cracks opening a direct path to metal. If there is a flake on top of this crack, there could be an extremely high local electric field creating electron emission into LXe. The emission can continue **indefinitely** until the flake changes or the field relaxes because the Ni/Fe/Cr wire metal is supplying electrons. One should remember that oxide thickness is only ~2 nm thick.

Some hot spots can also be initiated by S2 signal. I wrote a toy Monte Carlo code to simulate hot spot creation. Figure 10 describes the result of this simulation. First, I assume a perfect wire with undamaged citric-oxide layer - see Fig.10a; It is assumed that S2 occurs at x=0, y=0 position (light not shown) and QE = 5 x $10^{-4}$, i.e., perfectly passivated s.s. wires with ~2nm thick citric-oxide film without any faults. Probability of detection is QE*$N_{photons/e-}$ *$P_{ext_gate}$*$P_{hit_gate}$, where QE is quantum efficiency of photoelectron production on gate wires, $N_{photons/e-}$ is number of photons created from each electron in 8mm-thick gaseous section of the extraction region, $P_{ext_gate}$ is probability that photoelectron created on gate wire reaches ER and $P_{hit_gate}$ is probability to hit gate wire. The x-y image was made with artificially perfect detector imaging individual photoelectrons. Then, 3 damaged spots were introduced. Figure 10b shows three damages spots of several radii (5.6, 3.6 and 2.5 mm) and magnification factors M (10, 50 and 100), where M is a multiplication factor to multiply nominal QE of undamaged wire. Figure 10c shows what happens if field enhancement increases the factor M 200x, for all three spots. It is clear that if the factor M can be increased up to 1000x by field enhancement, hot spots will be very visible. They will produce large photonic activity which in turn will produce electron trains from stored charge in the oxide layer.

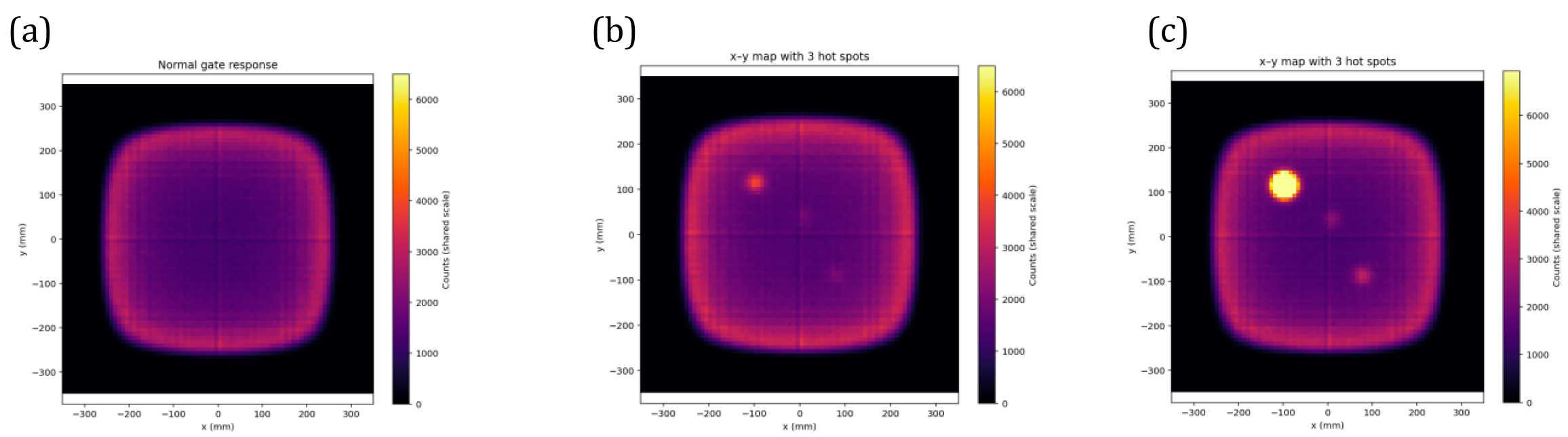

**Fig. 10** Imaging of photoelectrons produced on gate wires by light from S2 in extraction region, located at x=0, y=0 (not imaged). (a) gate wire oxide not damaged by thermal effects, (b) three hot spots caused by damages spots of several radii (5.6, 3.6 and 2.5 mm) and magnification factors M (10, 50 and 100), (c) the same as (b) but magnification factor M for all three spots is 200x.

## c) Ion-induced emission – Malter-like effect

Figure 11 shows an example of an ion landing on oxide, which is fully charged from previos VUV activity. There will be a huge electric field between ion and electron 1nm away; vacuum electric field between two charges is $E_{vacuum} \approx (k \cdot e)/d^2$, electric field with oxide is $E_{film} \approx E_{vacuum}/\varepsilon_r$ ~$1.44 x 10^9/\varepsilon_r$ V/m ~**720 kV/cm** for d = 1 nm and relative permittivity $\varepsilon_r$ ~ 20 for $Cr_2O_3/Cr(OH)_3/Cr_2O_{3-x}$ mix. Such field will likely trigger Malter-like electron emission of electrons from oxide capacitor.

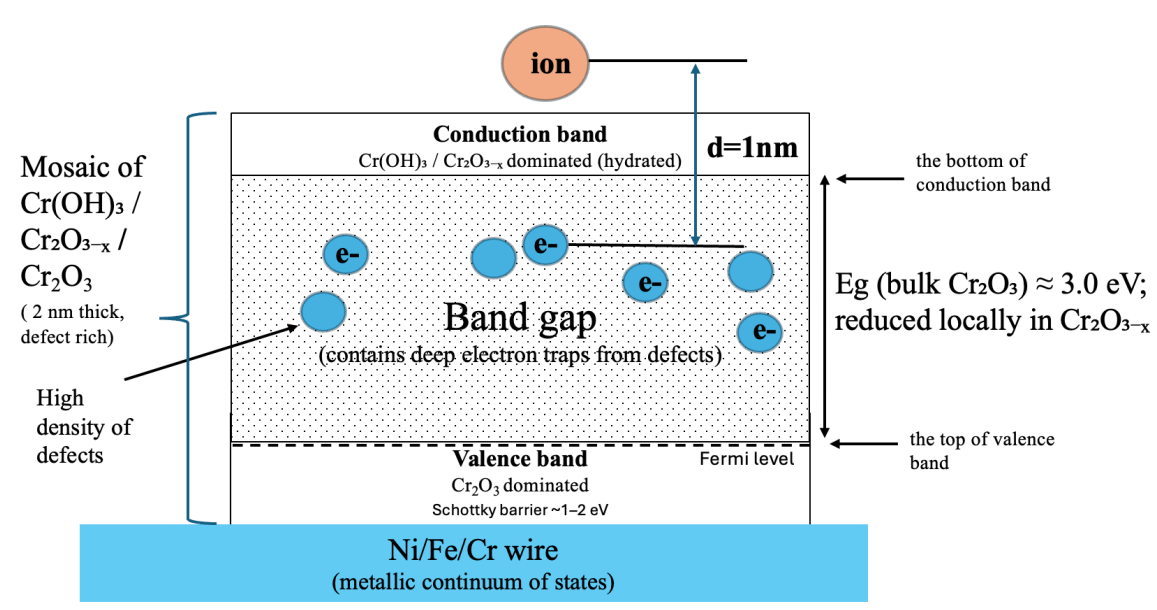

**Fig. 11** An ion sitting on oxide about 1 nm away from electron in band gap will create electric field of 720 kV/cm.

A classical Malter effect is triggered by positive feedback between anode gain and cathode covered by insulating layer – see example in next chapter. This is not the case for LZ TPC, which carefully chooses anode-grid voltages to prevent any gas charge gain. To prove this, I calculate charge gain in LZ as a function of voltage as shown on Fig.12; I used the Magboltz program [12]. Since extraction region has electric field of ~6.2 kV/cm in gaseous part, I conclude that LZ does not have any electron charge gain, unless there is some local wire fault.

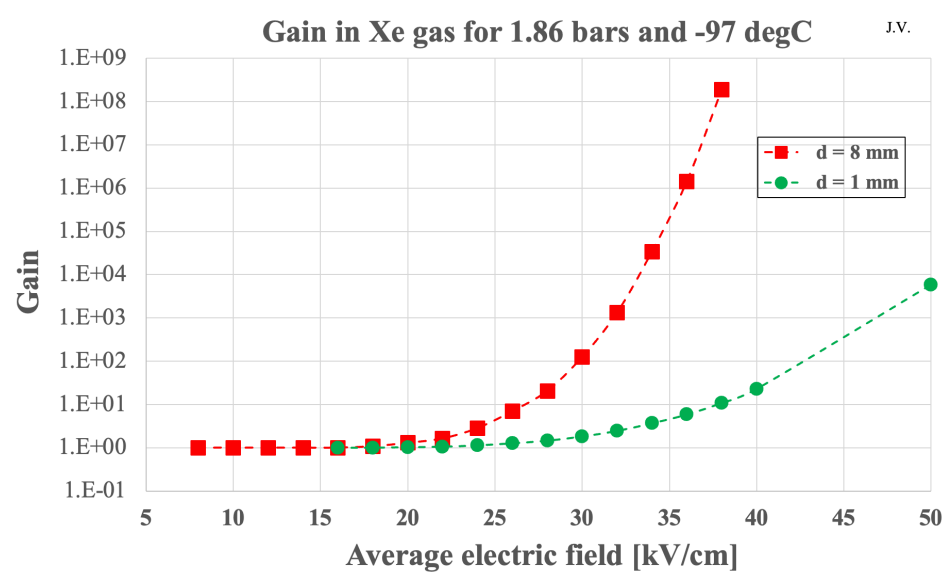

**Fig. 12** Magboltz program shows that charge gain starts at electric field of ~22 kV/cm, if we assume path of 8mm, and at ~28 kV/cm, if we assume path of 1 mm. LZ runs well below these numbers in gaseous section of extraction region, assuming that there are no defects in wire planes.

## d) Ion sources in LXe TPC

A typical muon passing through LZ TPC may deposit up to ~7x$10^7$ of electron-ion pairs. There are perhaps 10-15 such muons per day at LZ depth. It is clear that these ions will trigger Malter-like electron emissions from cathode wires, if wire oxide capacitor was charged up previously by VUV light. The same applies for alpha background and any calibration producing positive ions. Because there is no charge gain in extraction region and thus no positive feedback, positive ions will not lead to standing current of the classical Malter effect, but they will contribute to the overall noise level. Due to resistivity of oxides on s.s. wires, these ions can last long time on wires after each muon or calibration.

## 7. Example of Malter effect in CRID

To illustrate the classical Malter effect, I show an example from CRID detector, which had 40 TPCs (each was 1.2 m long) with quartz windows providing a total photosensitive area was more than ~15m$^2$, each TPC had a single electron wire detector operated at an average gain of 2x$10^5$. Photosensitive gas was TMAE. Figures 13a&b show a very localized train of electrons. This effect was triggered by excessive rate UV fiber calibration during the initial running. The effect stopped after we reduced the calibration rate [13]. The effect could be easily reproduced in the lab using strong UV or $Fe^{55}$ sources. During SLD operation the Malter effect was never observed, although we were cautious and set a trip threshold for each detector to 300 nA, and we could monitor each detector current with nA precision. A detailed recipe how to handle the Malter effect was described in [14], but the most important point was to pay attention to single electron trains. CRID field cage was made using Cu-Be wires and never suffered from the field emission, which would be catastrophic for such a detector, because of its photosensitivity.

(a)

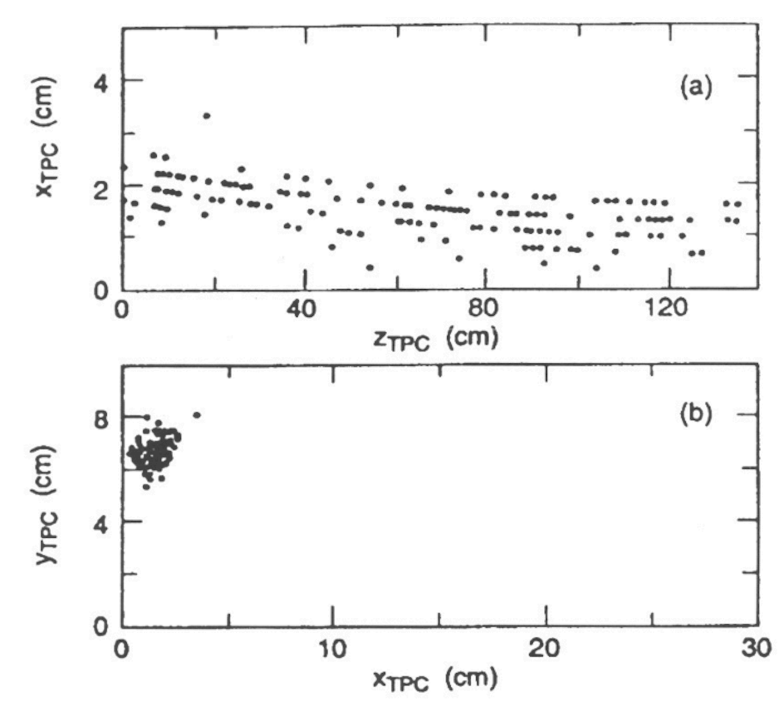

(b)

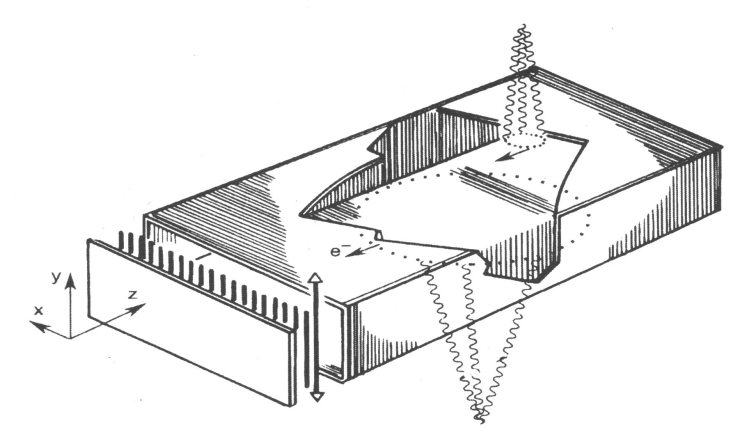

**Fig. 13** Malter effect observed in CRID TPC. It was positive feedback between anode and cathode, triggered by excessive fiber optics calibration rate. It stopped when the rate was reduced. (a) $x_{TPC}$-$z_{TPC}$ view shows a long train of single electron pulses along the entire length of TPC length; $x_{TPC}$-$y_{TPC}$ view of the same event. (b) Concept of CRID TPC (there were 40 of them). TPC top and bottom faces were made of quartz. Photosensitivity was achieved by TMAE gas addition, making it very sensitive to any electron and light emissions. We never saw a light emission from Cu-Be field cage wires.

## 8. Other contributions to photonic tail

We mention two other possible contributions to photonic tails, smaller in my opinion.

**Quartz scintillation** in does exist can be excited by passing particles or VUV light. This was measured by small test, which used beta-source shining into small test cell with PMT and SiPM, both equipped with quartz windows [15]. The test used beta source shining into a small test chamber cooled to −99 °C. Electrons entered quartz window of R8778 PMT and Hamamatsu S13371 SiPM, produced Cherenkov light, which in turn produced scintillation. The test measured secondary light with a tail lasting up to a millisecond. The conclusion was that the majority of it is due to fluorescence of quartz photosensor windows, following exposure to UV photons from xenon scintillation.

My comment is that this test does not explain the electron trains formation and hot spots in LZ, it cannot explain a second lasting tails, but it may contribute some fraction of it.

One should note that DIRC group has also measured scintillation background in DIRC fused silica bar when muon passes through. It was found that scintillation does exist, but it is negligible (<1%) compared to total number of photons muon produces passing through bar, i.e., Cherenkov light and delta rays [16].

**Photocathode** might also produce late pulses. A simple test with Photonics XP-2020 revealed that when it is exposed to daylight and then placed into dark box, it takes 100 minutes to bring its noise to low level. In the following test the same PMT was exposed to UV light pulse from Hamamatsu Xe-flash lamp and it was observed that there are late pulses lasting for ~10-15 milliseconds.[2] What I am suggesting that LZ PMT at LXe temperature should be exposed to 178 nm light pulses and determine possible late pulses.

I argue that photocathode can produce train of late pulses, however, it is only a contributor but not the primary driver of the problem discussed in this paper.

## Conclusion

I believe that this paper has provided explanation of long tails in S2 pulses. It is driven by long discharge time constant of the oxide capacitor on surface of s.s. wires. This discharge is in a form of observed long electron trains.

The paper also provided explanation of hot spots, which are created by cracking and flaking of oxide layer on s.s. wires during cooling to LXe temperature.

Choice of wires in LXe dual-phase TPCs is not a trivial problem. I recommend **abandoning passivation of s.s. grid wires**, and consider **gold plated Cu-Be wires** (One should realize that copper oxide might have similar problems as chromium oxide at LXe temperature.),[3] or gold plated s.s. wires. In both cases one has to verify the choice by many tests.

A positive ion can trigger a discharge of oxide capacitor, thus creating Malter-like electron emission and faking S2.

[2] Hamamatsu estimates that late light should last only ~ 100 microseconds.

Stainless steel electrodes were successfully used by single-phase LXe or LAR experiments such as ICARUS or DUNE, but they are not after detection of single photons and single electrons. But the mechanism described in this paper will contribute to noise.

## Acknowledgement

I would like to thank SLAC LZ group for providing information about the LZ operating experience.

[3] Webb telescope has gold plated beryllium mirrors.

# Appendix 1

## Table 2 - QE of various metals under various condition

| **Material / surface** | **QE ($e^-$ / photon)** | **Conditions** | **Source / Comment** |
|---|---|---|---|
| **Cu (clean, UHV)** | $\sim 10^{-4}$ – $3\times10^{-3}$ | 7 eV photons, oxide removed by Ar-ion sputtering | R. D. Young *Phys. Rev.* 113 (1959) 110; A. H. Sommer *Photoemissive Materials* (1968), ch. 3 |
| **Cu (air-exposed)** | $10^{-6}$ – $10^{-5}$ | Oxidized Cu ($Cu_2O$/CuO layer present) | S. Krolikowski & J. Spicer *Phys. Rev.* 185 (1969) 882 ("oxidized copper QE drop ≈ 100×") |
| **Al (clean)** | $\sim 10^{-3}$ | near threshold, oxide removed | A. H. Sommer (1968); D. Himpsel *Surf. Sci.* 115 (1982) L159 |
| **Al (oxidized)** | $10^{-6}$ – $10^{-5}$ | thin $Al_2O_3$ layer (~2 nm) | P. Yeh *J. Appl. Phys.* 57 (1985) 1689 (10.1063/1.334177) |
| **$Cr_2O_3$-covered SS304 (passivated)** | $(5 – 10)\times10^{-5}$ | Measured @ 178 nm, LAr/LXe temp | T. Haruyama *Nucl. Instrum. Meth.* A 819 (2016) 154; M. Kazama *arXiv:2503.14819* (2025) |
| **Lightly oxidized SS304 (air-aged)** | $3\times10^{-4}$ | 178 nm in vacuum | M. Kazama *arXiv:2503.14819* (2025); sample already had thin oxide |
| **Citric-acid passivated SS304 ($Cr(OH)_3$ / $Cr_2O_{3-x}$ / $Cr_2O_3$)** | $1\times10^{-4}$ – $5\times10^{-4}$ (typ.); up to $\sim 10^{-3}$ if defective or hydrated | Hydrated mixed oxide (~1–2 nm), band gap ≈ 2.3 eV, work function ≈ 4.8 eV; citric-acid passivation used by LZ | A. K. Shukla *Corros. Sci.* 52 (2010) 102; J. T. Kim & J. R. Scully *Corros. Sci.* 43 (2001) 485; M. J. Walzak *Appl. Surf. Sci.* 137 (1999) 143; T. Ohno *J. Mater. Res.* 18 (2003) 2109 |
| **Bare Fe–Cr–Ni metal (oxide-free patches)** | $10^{-3}$ – $10^{-2}$ | Inferred metallic regions w/out oxide, field enhancement at grain edges | Extrapolated from A. H. Sommer (1968) and G. P. Weisberg *J. Appl. Phys.* 34 (1963) 3330 [values compiled in Sommer] |